\documentstyle[11pt]{article}
\begin{document}

\title{\textbf{QCD ghost $f(T)$-gravity model}}

\author{K. Karami$^{1}$\thanks{KKarami@uok.ac.ir} ,
A. Abdolmaleki$^{1}$, S. Asadzadeh$^{1}$, Z. Safari$^{2}$\\
$^{1}$\small{Department of Physics, University of Kurdistan,
Pasdaran St., Sanandaj, Iran}\\$^{2}$\small{Research Institute for
Astronomy and Astrophysics of Maragha (RIAAM), Maragha, Iran}}

\maketitle

\begin{abstract}
Within the framework of modified teleparallel gravity, we
reconstruct a $f(T)$ model corresponding to the QCD ghost dark
energy scenario. For a spatially flat FRW universe containing only
the pressureless matter, we obtain the time evolution of the torsion
scalar $T$ (or the Hubble parameter). Then, we calculate the
effective torsion equation of state parameter of the QCD ghost
$f(T)$-gravity model as well as the deceleration parameter of the
universe. Furthermore, we fit the model parameters by using the
latest observational data including SNeIa, CMB and BAO data. We also
check the viability of our model using a cosmographic analysis
approach. Moreover, we investigate the validity of the generalized
second law (GSL) of gravitational thermodynamics for our model.
Finally, we point out the growth rate of matter density
perturbation. We conclude that in QCD ghost $f(T)$-gravity model,
the universe begins a matter dominated phase and approaches a de
Sitter regime at late times, as expected. Also this model is
consistent with current data, passes the cosmographic test,
satisfies the GSL and fits the data of the growth factor well as the
$\Lambda$CDM model.
\end{abstract}


\clearpage

\section{Introduction}

Astronomical data from the type Ia supernovae (SNeIa), cosmic
microwave background (CMB) and baryon acoustic oscillation (BAO),
etc., have revealed that the universe is undergoing an accelerating
expansion \cite{Riess}. This unexpected observed phenomenon poses
one of the most puzzling problems in cosmology today. There are two
representative approaches to explain this behavior. One is to
introduce some unknown matters with negative pressure called ``dark
energy'' (DE) in the framework of Einstein's general relativity (for
reviews on DE, see e.g. \cite{Padmanabhan}). The other approach for
describing the accelerated expansion of the universe is to modify
the gravitational theory called ``dark gravity'' (see e.g.
\cite{Tsujikawa,CapReview} for a review on modified gravity).

More recently, a new interesting DE model called ghost DE (GDE) has
been motivated from the Veneziano ghost of choromodynamics (QCD)
\cite{Urban}. In this proposal, it is claimed that the vacuum energy
arises from the contribution of the ghost fields which are supposed
to be present in the low energy effective theory of QCD. The
Veneziano ghost is required to exist for the resolution of the
$U(1)_A$ problem, but are completely decoupled from the physical
sector. The above claim is that the ghosts are decoupled from the
physical states and make no contribution in the usual Minkowski
space-time, but make a small energy density contribution to the
vacuum energy due to the off-set of the cancellation of their
contribution in the space-time with nontrivial topology or
time-dependent background such as our Friedmann-Robertson-Walker
(FRW) universe. This ghost gives the vacuum energy density
proportional to $\sim\Lambda^3_{\rm QCD}H$, where $H$ is the Hubble
parameter and $\Lambda_{\rm QCD}\sim100$ MeV is the QCD mass scale
\cite{Witten}. This small contribution can play an important role in
the evolutionary behavior of the universe. For instance, taking
$H\sim10^{-33}$ eV at the present, $\Lambda^3_{\rm QCD}H$ gives the
right order of observed magnitude of the DE density. This
coincidence is remarkable and implies that the GDE model gets rid of
fine tuning problem \cite{Urban}. In addition, the appearance of the
QCD scale could be relevant for a solution to the cosmic coincidence
problem, as it may be the scale at which dark matter (DM) forms
\cite{Forbes}. It is worth to note that the GDE model does not
violate unitarity, causality, gauge invariance and other important
features of renormalizable quantum field theory, as advocated in
\cite{Zhitnitsky1}. This new kind of DE model has got a lot of
enthusiasm recently in the literature
\cite{Cai,Sheykhi1,Sheykhi2,KKaramiNew}.

In the framework of modified gravity, the underlying philosophy of
extended theories of gravity is that general relativity (GR) should
be seen as a particular case of a more general effective theory
coming from fundamental principles. The common property of all these
approaches is that the DE effects can be associated to an evolving
equation of state (EoS). The basic idea lies on the fact that the
standard Einstein-Hilbert action is modified by additional degrees
of freedom, spanning from further curvature invariant corrections,
to scalar fields and Lorentz violating terms. The first need of
``corrections'' to GR emerges when quantum field theory is
formulated on curved space-time. Normalization and regularization
processes lead to non-minimal couplings and higher-order corrections
in curvature invariants \cite{CapReview}. Here, we limit our
attention to investigate the so-called $f(T)$ theories
\cite{Ferraro,bengochea}, which represent a class of models that
take into account the effects due to the torsion. Indeed, $f(T)$
theory is based on the old idea of the ``teleparallel'' equivalent
of GR (TEGR) \cite{Einstein}, which, instead of using the curvature
defined via the Levi-Civita connection, uses the Weitzenb\"{o}ck
connection that has no curvature but only torsion. In fact, this
approach was taken by Einstein \cite{Einstein} in an attempt of
unifying gravity and electromagnetism. TEGR is closely related to
standard GR, differing only in terms involving total derivatives in
the action, i.e. boundary terms \cite{Moller}. $f(T)$-gravity is a
modification of the teleparallel gravity in which the teleparallel
Lagrangian density described by the torsion scalar $T$ has been
promoted to a function of $T$. This concept is similar to the idea
of $f(R)$-gravity. However, in comparison with $f(R)$-gravity, whose
fourth-order equations may lead to pathologies, $f(T)$-gravity has
the significant advantage of possessing second-order field equations
\cite{WufT}. This feature has led to a rapidly increasing interest
in the literature. Models based on $f(T)$-gravity can provide an
alternative to inflation \cite{Ferraro}. It was also found that
$f(T)$ theory can explain the observed acceleration of the universe
\cite{bengochea}. Some viable phenomenological $f(T)$ models were
proposed by \cite{Linder}. Observational constraints were considered
in \cite{Wu}. A reconstruction of the $f(T)$ theory from the
background expansion history and the $f(T)$ theory driven by scalar
fields were studied in \cite{Myrzakulov}. It was shown that $f(T)$
theories are not dynamically equivalent to teleparallel action plus
a scalar field via conformal transformation \cite{Yang}.
Cosmological perturbations and growth factor of matter perturbations
in $f(T)$-gravity were investigated in \cite{Saridakis}. In
\cite{Meng}, Birkhoff's theorem in $f(T)$-gravity was studied.
Static solutions with spherical symmetry in $f(T)$ theories were
discussed in \cite{Wang}. In \cite{CapCosmo,CapCosmoRevised}, the
cosmic expansion was studied by using cosmography. Thermodynamical
description of $f(T)$-gravity was studied in \cite{Miao,KA}.

In the present work, our aim is to reconstruct a $f(T)$-gravity
model without resorting to any additional DE, that is, considering
that the ghost DE is effectively described by the modification of
the gravity with respect to the teleparallel gravity. To do so, in
section 2, we briefly review the $f(T)$-gravity in a spatially flat
FRW universe filled only with the pressureless matter. In section 3,
we reconstruct a $f(T)$ model according to the evolution of GDE
density. In section 4, we fit this model and give the constraints on
model parameters, with current observational data including SNeIa,
CMB and BAO data. In section 5, we check the viability of our model
using the cosmographic analysis method. In section 6, the validity
of the generalized second law of gravitational thermodynamics for
our $f(T)$ model is examined. In section 7, we study the growth of
structure formation in our model. Section 8 is devoted to
conclusions.

\section{$f(T)$-gravity}

The action of $f(T)$-gravity is given by \cite{Ferraro,bengochea}
\begin{equation}
I =\frac{1}{2k^2}\int {\rm d}^4
x~e~\Big[f(T)+L_m\Big],\label{action}
\end{equation}
where $k^2=8\pi G$, $e={\rm det}(e^i_{\mu})$ and $e^i_{\mu}$ is the
vierbein field which is used as a dynamical object in the
teleparallel gravity. Also $T$ and $L_m$ are the torsion scalar and
the Lagrangian density of the matter inside the universe,
respectively.

Taking the variation of the action (\ref{action}) with respect to
the vierbein $e^i_{\mu}$, the modified Friedmann equations in the
spatially flat FRW universe can be obtained in the standard forms
\begin{equation}
\frac{3}{k^2}H^2=\rho_m+\rho_T,\label{fT11}
\end{equation}
\begin{equation}
\frac{1}{k^2}(2\dot{H}+3H^2)=-(p_m+p_T),\label{fT22}
\end{equation}
where
\begin{equation}
\rho_T=\frac{1}{2k^2}(2Tf_T-f-T),\label{roT}
\end{equation}
\begin{equation}
p_T=-\frac{1}{2k^2}[-8\dot{H}Tf_{TT}+(2T-4\dot{H})f_T-f+4\dot{H}-T],\label{pT}
\end{equation}
and
\begin{equation}
T=-6H^2.\label{T}
\end{equation}
Here, $H=\dot{a}/a$ is the Hubble parameter and the subscript $T$
denotes a derivative with respect to the torsion scalar $T$. Also
$\rho_m$ and $p_m$ are the energy density and pressure of the matter
inside the universe, respectively. Furthermore, $\rho_T$ and $p_T$
are the torsion contributions to the energy density and pressure,
respectively. Note that in the case of $f(T)=T$, Eqs. (\ref{roT})
and (\ref{pT}) give $\rho_T=0$ and $p_T=0$. Then Eqs. (\ref{fT11})
and (\ref{fT22}) recover the usual Friedmann equations in the
teleparallel gravity.

The energy conservation equations are still given by
\begin{equation}
\dot{\rho}_m+3H(\rho_m+p_m)=0,\label{conteq}
\end{equation}
\begin{equation}
\dot{\rho}_T+3H(\rho_T+p_T)=0.\label{EfT}
\end{equation}
The effective torsion EoS parameter is defined as \cite{WufT,KA}
\begin{equation}
\omega_T=\frac{p_T}{\rho_T}=-1-\frac{\dot{T}}{3H}\left(\frac{2Tf_{TT}+f_T-1}{2Tf_T-f-T}\right).\label{omegaT}
\end{equation}
In the de Sitter universe, i.e. $\dot{H}=0=\dot{T}$, Eq.
(\ref{omegaT}) yields $\omega_T=-1$ which behaves like the
$\Lambda$CDM model.

With the help of Eqs. (\ref{fT11}), (\ref{roT}) and (\ref{T}) one
can get
\begin{equation}
\rho_m=\frac{1}{16\pi G}(f-2Tf_T).\label{rhom}
\end{equation}
For the pressureless matter, i.e. $p_m=0$, from Eqs. (\ref{fT11}) to
(\ref{pT}) one can obtain
\begin{equation}
\dot{H}=-\frac{4\pi G\rho_m}{f_T+2Tf_{TT}}.\label{Hdot}
\end{equation}
Inserting Eq. (\ref{rhom}) into (\ref{Hdot}) and using
$\dot{T}=-12H\dot{H}$ gives
\begin{equation}
\dot{T}=3H\left(\frac{f-2Tf_{T}}{f_T+2Tf_{TT}}\right).\label{Tdot}
\end{equation}
Using the above relation, the effective EoS parameter (\ref{omegaT})
yields
\begin{equation}
\omega_T=-\frac{f/T-f_T+2Tf_{TT}}{(f_T+2Tf_{TT})(f/T-2f_T+1)}.\label{omegaT2}
\end{equation}
Here, we also calculate the deceleration parameter
\begin{equation}
q=-1-\frac{\dot{H}}{H^2},\label{q1}
\end{equation}
which can be compared with the observations. Using Eqs. (\ref{T})
and (\ref{Tdot}) the deceleration parameter (\ref{q1}) leads to
\begin{equation}
q=2\left(\frac{f_T-Tf_{TT}-\frac{3f}{4T}}{f_T+2Tf_{TT}}\right).\label{q2}
\end{equation}
For $f(T)=T$, from Eq. (\ref{q2}) we have $q=0.5$ which corresponds
to the matter dominated universe.

\section{Ghost $f(T)$-gravity model}

The dark torsion contribution in $f(T)$-gravity can justify the
observed acceleration of the universe without resorting to the DE.
This motivates us to reconstruct a $f(T)$-gravity model according to
the GDE model. The GDE density, which comes from the Veneziano ghost
of QCD, is proportional to the Hubble parameter \cite{Urban,Cai}
\begin{equation}\label{GDE1}
\rho_D=\alpha H,
\end{equation}
where $\alpha$ is a constant with dimension $[{\rm energy}]^3$, and
roughly of order of $\Lambda_{\rm QCD}^3$ where $\Lambda_{\rm
QCD}\sim100$ MeV is the QCD mass scale.

With the help of Eq. (\ref{T}) one can rewrite (\ref{GDE1}) as
\begin{equation}
\rho_D=\alpha\left(-\frac{T}{6}\right)^{1/2}.\label{GDE2}
\end{equation}
Equating (\ref{roT}) and (\ref{GDE2}), i.e. $\rho_T=\rho_D$, we
obtain the following differential equation
\begin{equation}
2Tf_T-f-T-\beta\sqrt{-T}=0,\label{diffEq}
\end{equation}
where
\begin{equation}
\beta=\frac{2k^2\alpha}{\sqrt6}.\label{beta}
\end{equation}
Solving Eq. (\ref{diffEq}) yields the $f(T)$-gravity corresponding
to the QCD ghost DE model as
\begin{equation}
f(T)=T+\sqrt{-T}\left(\epsilon+\frac{\beta}{2}\ln(-T)\right),\label{f(T)GDE1}
\end{equation}
where $\epsilon$ is an integration constant that can be determined
from a boundary condition. Following \cite{CapCosmo} to recover the
present day value of Newtonian gravitational constant we need to
have
\begin{equation}
f_T(T_0)=1,\label{f_T}
\end{equation}
where $T_0=-6H_0^2$ is the torsion scalar at the present time.
Applying the above boundary condition to the solution
(\ref{f(T)GDE1}) one can obtain
\begin{equation}
\epsilon=-\beta\left(1+\frac{1}{2}\ln(-T_0)\right).\label{ep}
\end{equation}
Substituting this into Eq. (\ref{f(T)GDE1}) gives
\begin{equation}
f(T)=T+\beta\sqrt{-T}\left[\frac{1}{2}\ln\left(\frac{T}{T_0}\right)-1\right].\label{f(T)GDE}
\end{equation}
Note that the parameter $\beta$ can be obtained by inserting Eq.
(\ref{f(T)GDE}) into the modified Friedmann equation (\ref{fT11}).
Solving the resulting equation for the present time gives
\begin{equation}
\beta=\sqrt{6}H_0(1-\Omega_{m_0}),\label{beta}
\end{equation}
where $\Omega_{m_0}=\frac{k^2\rho_{m_0}}{3H_0^2}$ is the
dimensionless matter energy density and the index 0 denotes the
value of a quantity at present.

\begin{figure}
\includegraphics{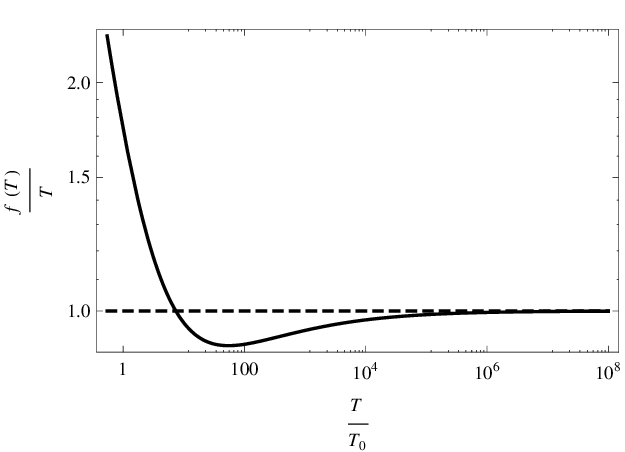}
      \vspace{7.0cm}
\caption[]{The evolution of QCD ghost $f(T)$-gravity model, Eq.
(\ref{f(T)GDE}), versus $T/T_0$. The dashed line denotes the model
$f(T)=T$ corresponding to the case of teleparallel gravity for
comparison. Auxiliary parameters are: $H_0=70.6~{\rm
km~s^{-1}~Mpc^{-1}}$ \cite{CapCosmo} and $\Omega_{m_0}=0.262$. For
these values one finds $\beta=\sqrt{6}H_0(1-\Omega_{m_0})=127.625$
and $f(T_0)=-51976.9$.}
         \label{f-fEinstein-T}
   \end{figure}

The evolution of the ghost $f(T)$-gravity model, Eq.
(\ref{f(T)GDE}), versus $T/T_0$ is shown in Fig.
\ref{f-fEinstein-T}, where we also plot $f(T)=T$ corresponding to
the case of teleparallel gravity for comparison. Figure
\ref{f-fEinstein-T} shows that the QCD ghost $f(T)$-gravity model
(\ref{f(T)GDE}) satisfies the condition
\begin{equation}
\lim_{|T|\rightarrow\infty}f/T\rightarrow 1,
\end{equation}
at high redshift which is compatible with the primordial
nucleosynthesis and CMB constraints \cite{WufT,Linder}.

\begin{figure}
\includegraphics{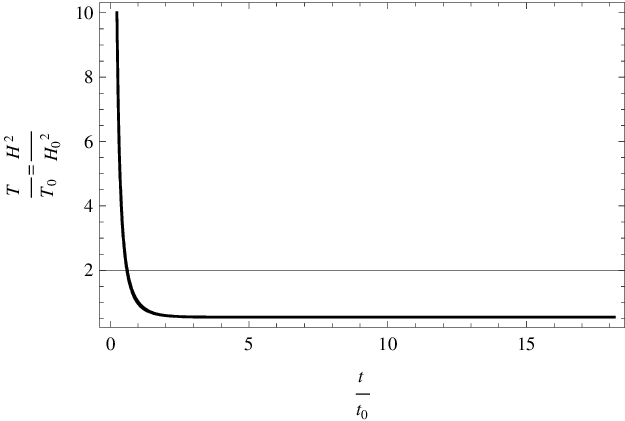}
      \vspace{7.0cm}
\caption[]{The evolution of the torsion scalar (or the squared
Hubble parameter), Eq. (\ref{t1}), versus $t/t_0$. Auxiliary
parameters as in Fig. \ref{f-fEinstein-T}.}
         \label{T-time}
   \end{figure}

Inserting Eq. (\ref{f(T)GDE}) into (\ref{Tdot}) and using
$H=(-T/6)^{1/2}$ yields
\begin{equation}
t-t_i=\frac{1}{\sqrt{6}}\int_{T_{\rm
i}}^{T}\left(\frac{2\sqrt{-T}-\beta}{\beta\sqrt{-T}+T}\right)\frac{{\rm
d}T}{T}.\label{t}
\end{equation}
Integrating the above relation analytically gives
\begin{equation}
t=\sqrt{\frac{2}{3}}\left[\frac{1}{\sqrt{-T}}
+\frac{1}{\beta}\ln\left(\frac{\sqrt{-T}}{\sqrt{-T}-\beta}\right)\right],~~~\sqrt{-T}\geq\beta,\label{t1}
\end{equation}
where at $T_i=-6H_i^2=-\infty$ we have $t_i=0$. Note that the
condition $\sqrt{-T}\geq\beta$ is necessary due to having a real
time. Using Eq. (\ref{beta}) the condition $\sqrt{-T}\geq\beta$ can
be rewritten as $T/T_0\geq(1-\Omega_{m_0})^2$. Figure \ref{T-time}
shows time evolution of the fractional torsion scalar
$T/T_0=(H/H_0)^2$ (or the fractional squared Hubble parameter). It
clears that $T/T_0$ (or $H^2/H_0^2$) decreases with increasing the
time. Figure \ref{T-time} illustrates that at early
($t/t_0\rightarrow 0$) and late ($t/t_0\rightarrow\infty$) times we
have $T/T_0\rightarrow\infty$ and
$T/T_0\rightarrow(1-\Omega_{m_0})^2=0.545$, respectively, where we
take $\Omega_{m_0}=0.262$ from the cosmological constraints (see
section 4).

It is worth to mention that in \cite{Myrzakulov}, to reconstruct a
$f(T)$-gravity according to a specific DE model, usually an ansatz
for the scale factor $a(t)$ or the Hubble parameter $H(t)$ is
assumed. Although this selection may
 justify the asymptotically behavior of the universe,
the obtained $f(T)$ model doesn't satisfy the full set of Eqs.
(\ref{fT11}), (\ref{fT22}), (\ref{conteq}) and (\ref{EfT}). Because
one cannot assume both a relation $a(t)$ (or $H(t)$) and a relation
$\rho_T=\rho_D(t)$, independently. Choosing one determines the other
through the Friedmann equations (\ref{fT11})-(\ref{fT22}) and so it
is inconsistent to choose both. Whereas in the present work, one can
determine the time evolution of the Hubble parameter
$H=(-T/6)^{1/2}$ with the help of Eq. (\ref{t1}), as plotted in Fig.
\ref{T-time}. Besides, our $f(T)$ model (\ref{f(T)GDE}) satisfies
the full set of field equations in $f(T)$-gravity.

\begin{figure}
\includegraphics{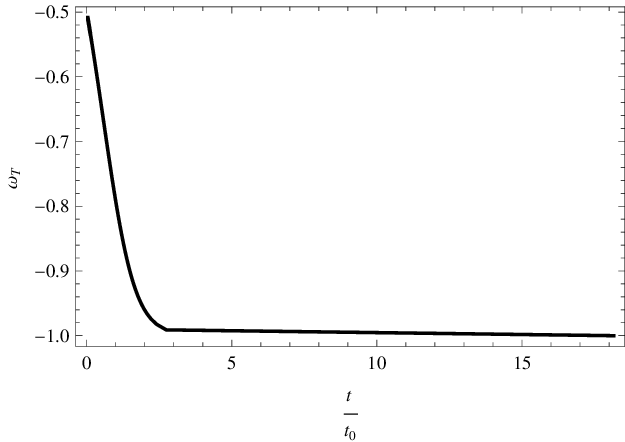}
      \vspace{7.0cm}
\caption[]{The effective torsion EoS parameter of the QCD ghost
$f(T)$-gravity model, Eq. (\ref{omegaGDE2}), versus $t/t_0$.
Auxiliary parameters as in Fig. \ref{f-fEinstein-T}.}
         \label{wT-time}
  \end{figure}

Inserting Eq. (\ref{f(T)GDE}) into (\ref{omegaT2}) gives the
effective torsion EoS parameter of the ghost $f(T)$-gravity model as
\begin{equation}
\omega_T=-\frac{T}{2T+\beta\sqrt{-T}}.\label{omegaGDE2}
\end{equation}
The time evolution of the EoS parameter (\ref{omegaGDE2}) is plotted
in Fig. \ref{wT-time}. It shows that at early time
($t/t_0\rightarrow 0$) we have $\omega_T\rightarrow-0.5$ and at late
time ($t/t_0\rightarrow\infty$) we get $\omega_T\rightarrow-1$ which
acts like the $\Lambda$CDM model. Also at present time we have
$\omega_{T_0}=-0.79$. Figure \ref{wT-time} clears that the torsion
EoS parameter of the ghost $f(T)$-gravity model behaves like
freezing quintessence DE \cite{Caldwell}. This result is in complete
agreement with that obtained for the GDE model in the Einstein
gravity \cite{Cai}.

\begin{figure}
\includegraphics{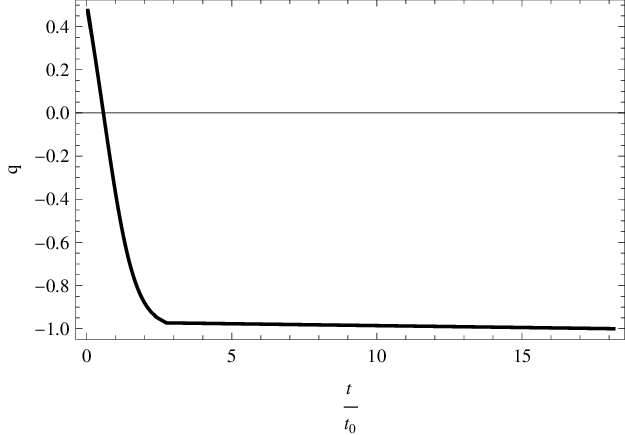}
      \vspace{7.0cm}
\caption[]{The deceleration parameter of the QCD ghost
$f(T)$-gravity model, Eq. (\ref{q3}), versus $t/t_0$. Auxiliary
parameters as in Fig. \ref{f-fEinstein-T}.}
         \label{q-time}
   \end{figure}

Inserting Eq. (\ref{f(T)GDE}) into (\ref{q2}) one can obtain the
deceleration parameter
\begin{equation}
q=\frac{\sqrt{-T}-2\beta}{2\sqrt{-T}-\beta}.\label{q3}
\end{equation}
Figure \ref{q-time} shows the time evolution of the deceleration
parameter (\ref{q3}). It clears that at early time
($t/t_0\rightarrow 0$) we have $q\rightarrow0.5$ which corresponds
to the matter dominated universe. Also at late time
($t/t_0\rightarrow\infty$) we get $q\rightarrow-1$ which behaves
like the de Sitter universe. Figure \ref{q-time} illustrates that at
the near past $t/t_0=0.57$ we have a cosmic deceleration $q>0$ to
acceleration $q<0$ transition which is compatible with the
observations \cite{Ishida}. Also at present time we get
$q_0\simeq-0.4$ which is in good agreement with the recent
observational results $-1.4\leq q_0\leq -0.3$ \cite{Ishida}.

\section{Cosmological constraints}

Here, we fit the free parameter $\Omega_{m_0}$ of the QCD ghost
$f(T)$-gravity model (\ref{f(T)GDE}) by using the recent
observational data including SNeIa, CMB shift and BAO data.

Since SNeIa behave as excellent standard candles, they can be used
to directly measure the expansion rate of the universe up to high
redshifts ($z\geq 1$) for comparison with the present rate.
Therefore, they provide direct information on the accelerating
universe and constrain the model. Recently, the Supernova Cosmology
Project (SCP) collaboration released the updated Union2.1
compilation which consists of 580 SNeIa \cite{union}. The Union2.1
compilation is the largest published and spectroscopically confirmed
SNeIa sample to date. The data points of the 580 Union2.1 SNeIa
compiled in \cite{union} are given in terms of the distance modulus
$\mu_{\rm obs}(z_{\rm i})$. From the SNeIa constraint, the best fit
value of the model parameters can be obtained by minimizing
\cite{Pietro,Nesseris}
\begin{equation}
\tilde{\chi}_{\rm SN}^{2}=A-\frac{B^{2}}{C},
\end{equation}
where
\begin{equation}
A=\sum_{\rm i=1}^{580}[\mu_{\rm obs}(z_{\rm i})-\mu_{\rm th}(z_{\rm
i})]^{2}/\sigma_{\rm i}^{2},
\end{equation}
\begin{equation}
B=\sum_{\rm i=1}^{580}[\mu_{\rm obs}(z_{\rm i})-\mu_{\rm th}(z_{\rm
i})]/\sigma_{\rm i}^{2},
\end{equation}
\begin{equation}
C=\sum_{\rm i=1}^{580}1/\sigma_{\rm i}^{2},
\end{equation}
and $\sigma_{\rm i}$ stands for the $1\sigma$ uncertainty associated
to the $i$th data point. Here, $\mu_{\rm th}(z)$ is the theoretical
distance module defined as \cite{Pietro,Nesseris}
\begin{equation}
\mu_{\rm th}(z)=5\log_{10}D_{\rm L}(z)+\mu_{0},
\end{equation}
where $\mu_{0}=42.38-5\log_{10}h$ and $h$ is the Hubble constant
$H_0$ in units of $100~\rm{km~s^{-1}~Mpc^{-1}}$. Also the
Hubble-free luminosity distance $D_{\rm L}(z)$ for the flat universe
is given by
\begin{equation}
 D_{\rm L}(z)=(1+z)\int_{0}^{z}\frac{dz'}{E(z';\textbf{p})},\label{DL-SNeIa}
 \end{equation}
with \textbf{p} the model parameters and
$E(z;\textbf{p})=H(z;\textbf{p})/H_0$.

Next, we add the data from the observation of the CMB. The CMB peak
from WMAP observations arises from acoustic oscillations of the
primeval plasma just before the universe becomes transparent. The
structure of the anisotropies of the CMB radiation depends on two
eras in cosmology including last scattering and today
\cite{YWang,Bond}. They can also be applied to limit the model
parameters. The $\chi^2$ from the CMB constraint is given by
\begin{equation}
\chi_{\rm CMB}^{2}=\frac{\left[R_{\rm obs}-R_{\rm
th}\right]^{2}}{\sigma_{R}^{2}}.
\end{equation}
Here, the theoretical value of CMB shift parameter, $R_{\rm th}$, is
defined as \cite{YWang,Bond}
\begin{equation}
R_{\rm th}=\sqrt{\Omega_{m_0}}\int_{0}^{z_{\rm
rec}}\frac{dz'}{E(z';\textbf{p})},\label{R-CMB}
\end{equation}
where $z_{\rm rec}\simeq1091.3$ is the redshift at the recombination
epoch, which has been updated in the 7-year WMAP (WMAP7) data
\cite{Komatsu}. Also the observational value of $R_{\rm obs}$ has
been updated to $1.725\pm0.018$ from the WMAP7 data \cite{Komatsu}.
The shift parameter $R$ relates the angular diameter distance to the
last scattering surface, the comoving size of the sound horizon at
$z_{\rm rec}$ and the angular scale of the first acoustic peak in
CMB power spectrum of temperature fluctuations \cite{YWang,Bond}.

Finally, we further add the data from the observation of the large
scale structure (LSS). Here, we use the distance parameter $A$ of
the measurement of the BAO peak in the distribution of Sloan Digital
Sky Survey (SDSS) luminous red galaxies \cite{Tegmark,Eisenstein},
which contains the main information of the observations of LSS.
Using the BAO data, one can minimize the $\chi_{\rm BAO}^{2}$
defined as \cite{Tegmark,Eisenstein},
\begin{equation}
\chi_{\rm BAO}^{2}=\frac{\left[A_{\rm obs}-A_{\rm
th}\right]^{2}}{\sigma_{A}^{2}},
\end{equation}
where
\begin{equation}
A_{\rm th}=\sqrt{\Omega_{m_0}}~E(z_{\rm
b};\textbf{p})^{-1/3}\left[\frac{1}{z_{\rm b}}\int_{0}^{z_{\rm
b}}\frac{dz'}{E(z';\textbf{p})}\right]^{2/3},\label{A-BAO}
\end{equation}
and $z_{\rm b}=0.35$ is the redshift of luminous red galaxies sample
of the SDSS. Also $A_{\rm obs}=0.469(n_{s}/0.98)^{-0.35}\pm 0.017$
is measured from the SDSS data \cite{Eisenstein} and the scalar
spectral index $n_{s}$ is taken to be 0.968, which has been updated
from the WMAP7 data \cite{Komatsu}.

\begin{figure}
\includegraphics{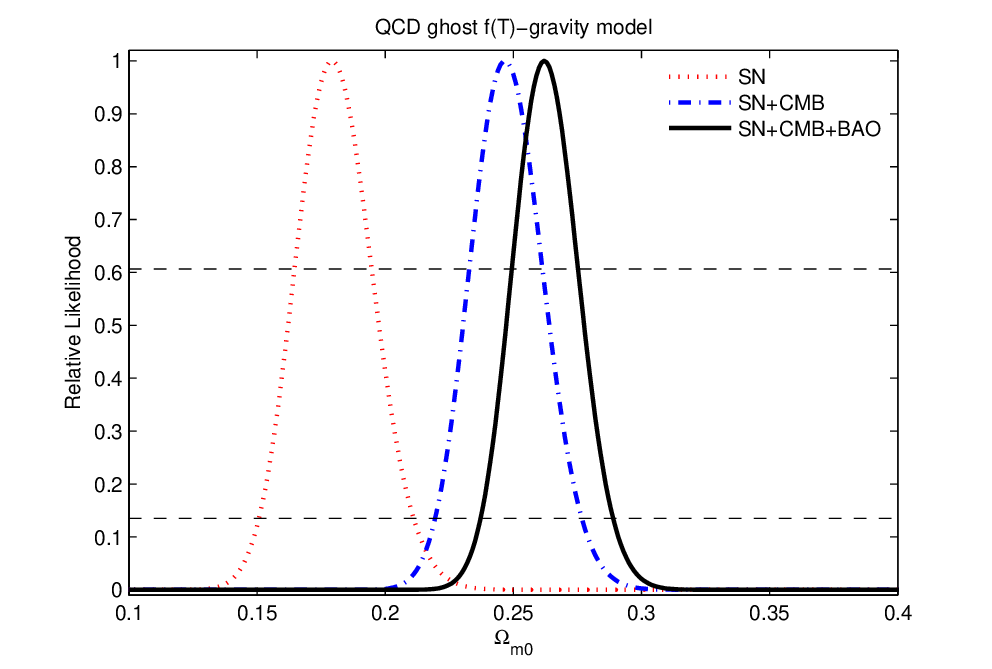}
      \vspace{4.7cm}
\caption{1D Likelihood for $\Omega_{m_0}$. Horizontal lines give the
bounds with $1\sigma$ and $2\sigma$ level of confidence.}
         \label{ghost_Likelihood}
   \end{figure}

Note that to fit the model parameters one needs to determine the
dimensionless Hubble parameter $E(z;\textbf{p})=H(z;\textbf{p})/H_0$
appeared in Eqs. (\ref{DL-SNeIa}), (\ref{R-CMB}) and (\ref{A-BAO}).
To do so, using Eqs. (\ref{fT11}), (\ref{roT}), (\ref{f(T)GDE}) and
(\ref{beta}) one can obtain the dimensionless Hubble parameter as
\begin{equation}
E(z;\textbf{p})=\frac{H(z;\textbf{p})}{H_0}=\frac{1}{2}(1-\Omega_{m_0})
+\left[\frac{1}{4}(1-\Omega_{m_0})^{2}+\Omega_{m_0}(1+z)^{3}\right]^{1/2}.\label{eqE}
\end{equation}
As the normalized likelihood function is defined by ${\mathcal
L}=e^{-(\chi^2_{\rm total}-\chi^2_{\rm min})/2}$ \cite{nesseris},
the best fit value of the model parameters follows from minimizing
the sum
\begin{equation}
\chi^2_{\rm total}=\tilde{\chi}^2_{\rm SN}+\chi^2_{\rm
CMB}+\chi^2_{\rm BAO}.
\end{equation}
The results are summarized in Table \ref{best-fit}, where we also
list the best fit value of the corresponding parameter of the
$\Lambda$CDM model for comparison. At 68.3\% and 95.4\% confidence
levels, we obtain the best fit value
$\Omega_{m_0}=0.262_{-0.013}^{+0.013}(1\sigma)_{-0.025}^{+0.027}(2\sigma)$
for the full data sets including SNeIa+CMB+BAO. The best fit value
obtained for $\Omega_{m_0}$ is slightly smaller than the
corresponding one in the $\Lambda$CDM model. The total $\chi^2$ of
the best fit value of the QCD ghost $f(T)$-gravity model is
$\chi_{\rm min}^{2}=606.567$ for the full data sets with degrees of
freedom (dof) $=582$. The reduced $\chi^2$ is $1.042$, which is
acceptable, but $\chi_{\rm min}^{2}$ is larger than the one for the
$\Lambda$CDM model, $\chi_{\Lambda\rm CDM}^{2}=562.531$, for the
same data sets. The relative (normalized) likelihood function
${\mathcal L}$ versus $\Omega_{m_0}$ is also shown in Figure
\ref{ghost_Likelihood}.

\section{Cosmographic analysis}

Here, using a cosmographic analysis approach introduced by
Capozziello et al. \cite{CapCosmo} we check the viability of our
model without the need of explicitly solving the field equations and
fitting the data. In this approach, the parameters of a given $f(T)$
model must be chosen in such a way that the model-independent
constraints on the cosmographic parameters ($h,q_0,j_0,s_0,l_0$),
listed in Table I in \cite{CapCosmo}, are satisfied. Here,
$h,q_0,j_0,s_0$ and $l_0$ are usually referred to as the Hubble
constant, deceleration, jerk, snap, and lerk parameters,
respectively. As we already obtained, imposing the condition
(\ref{f_T}) on our model (\ref{f(T)GDE1}) yields Eq. (\ref{ep}).
This yields the $f_i=f^{(i)}(T_0)/(6H_0^2)^{-(i-1)}$ values, given
by Eqs. (4.23)-(4.26) in Capozziello et al. \cite{CapCosmo}, for
$i=(2,3,4,5)$ where $f^{(i)}(T)={\rm d}^if/{\rm d}T^i$ to be
expressed as function of $\beta$ only when we fix
$\Omega_{m_0}=0.1329/h^2$ from the WMAP7 data. Following
\cite{CapCosmo} for each $f_2$ value of the sample obtained above
from the cosmographic parameters analysis, we solve
$\hat{f}_2(\beta)=f_2$. Then, we estimate the theoretically expected
values for the other derivatives $(f_3,f_4,f_5)$. The median and
68\% and 95\% confidence ranges are obtained as
\begin{eqnarray}
&&f_3=0.298_{-0.336-0.789}^{+0.150+0.197}\nonumber\\
&&f_4=0.857_{-0.966-2.269}^{+0.433+0.567}\nonumber\\
&&f_5=3.281_{-3.696-8.682}^{+1.658+2.172}.
\end{eqnarray}
Now we compare the above results with the model-independent
constraints on the $f_i$ values given in Table II in
\cite{CapCosmo}. Following \cite{CapCosmo} we use only the 68\%
confidence level which we compare the above constraints to. Our
comparison shows that the values of $(f_3,f_4,f_5)$ take place in
the 68\% confidence level in Table II in \cite{CapCosmo}. Therefore,
we conclude that QCD ghost $f(T)$-gravity model (\ref{f(T)GDE}) is
favored by the observational data.

\section{Generalized second law of thermodynamics}

\begin{figure}
\includegraphics{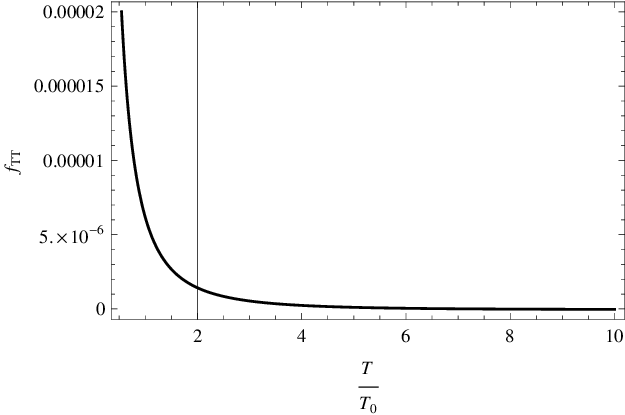}
      \vspace{7.0cm}
\caption[]{$f_{TT}$ versus $T/T_0$ for the QCD ghost $f(T)$-gravity
model (\ref{f(T)GDE}). Auxiliary parameters as in Fig.
\ref{f-fEinstein-T}.}
         \label{fTT-T}
   \end{figure}

Here, we examine the validity of the generalized second law (GSL) of
gravitational thermodynamics for our model. According to the GSL,
entropy of the matter inside the horizon beside the entropy
associated with the surface of horizon should not decrease during
the time \cite{Cai05}. Karami and Abdolmaleki \cite{KA} showed that
within the framework of $f(T)$-gravity, the GSL for a spatially flat
FRW universe enclosed by the dynamical apparent (Hubble) horizon
$\tilde{r}_{\rm A}=H^{-1}$ and containing only the pressureless
matter is given by
\begin{equation}
T_{\rm A}\dot{S}_{\rm
tot}=\frac{9}{8G}\left(\frac{f-2Tf_T}{f_T+2Tf_{TT}}\right)\left[
4f_{TT}+\left(\frac{f-2Tf_T}{f_T+2Tf_{TT}}\right)\left(\frac{f_T+5Tf_{TT}}{T^2}\right)\right]
,\label{TASdot}
\end{equation}
where $S_{\rm tot}=S_m+S_{\rm A}$ is the total entropy due to
different contributions of the matter and the horizon. Here, $T_{\rm
A}$ is the Hawking temperature on the apparent horizon
$\tilde{r}_{\rm A}$. Note that the horizon entropy in $f(T)$-gravity
given by
\begin{equation}
S_A=\frac{Af_T}{4G},\label{SA}
\end{equation}
where $A=4\pi \tilde{r}_{\rm A}^2$, is valid only when $f_{TT}$ is
small \cite{Miao}. To check this, we plot $f_{TT}$ versus $T/T_0$
for our model (\ref{f(T)GDE}) in Fig. \ref{fTT-T}. Figure shows that
the $f_{TT}$ is very small throughout history of the universe. This
confirms the validity of Eq. (\ref{SA}) for our model.

\begin{figure}
\includegraphics{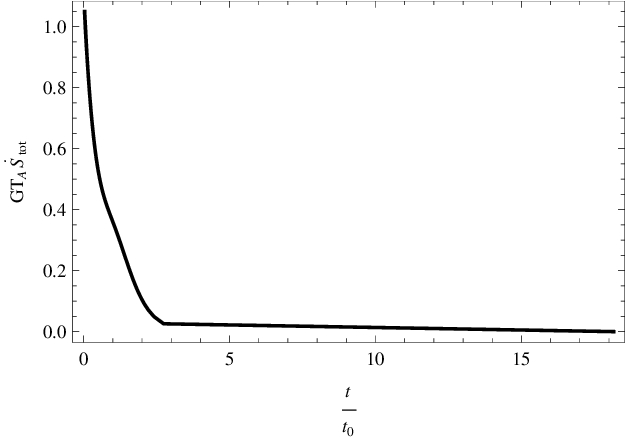}
      \vspace{7.0cm}
\caption[]{The variation of the GSL, Eq. (\ref{TSdotGDE}), versus
$t/t_0$ for model (\ref{f(T)GDE}). Auxiliary parameters as in Fig.
\ref{f-fEinstein-T}.}
         \label{GSL-time}
   \end{figure}

Now we examine the validity of the GSL, Eq. (\ref{TASdot}), for the
QCD ghost $f(T)$-gravity model (\ref{f(T)GDE}). The GSL reads
\begin{equation} GT_{\rm A}\dot{S}_{\rm
tot}=\frac{9}{16}\left[\frac{2\Big(4(-T)^{3/2}+9\beta
T+(8\sqrt{-T}-3\beta)\beta^2\Big)+\beta(T+\beta^2)\ln(T/T_0)}
{\sqrt{-T}~(\beta-2\sqrt{-T})^2}\right].\label{TSdotGDE}
\end{equation}
The variation of the GSL (\ref{TSdotGDE}) versus $t/t_0$ is plotted
in Fig. \ref{GSL-time}. Figure shows that the GSL for our model is
satisfied throughout history of the universe.

\section{Growth rate of matter density perturbation}
Here, we study the growth rate of matter density perturbation in QCD
ghost $f(T)$-gravity model. The origin of structure in the universe
is seeded by the small quantum fluctuations generated at the
inflationary epoch. These small perturbations over time grew to
become all of the structure we observe. Once the universe becomes
matter dominated primeval density inhomogeneities ($\delta\rho_{\rm
m}/\rho_{\rm m}\sim 10^{-5}$) are amplified by gravity and grow into
the structure we see today \cite{Rui Zheng}. The evolution equation
for the matter density contrast $\delta_{m}=\delta\rho_{m}/\rho_{m}$
in $f(T)$-gravity is given by \cite{Rui Zheng}
\begin{equation}
\ddot{\delta}_{m}+2H\dot{\delta}_{m}-4{\pi}G_{\rm
eff}\rho_{m}\delta_{m}=0,\label{eqdelta1}
\end{equation}
where $G_{\rm eff}$ is the effective Newton's constant which is
related to $G$ by $G_{\rm eff}=\frac{G}{f_T}$. In the case of
$f(T)=T$, we have $G_{\rm eff}=G$ and Eq. (\ref{eqdelta1}) recovers
the corresponding linear matter perturbation equation in TEGR.

Note that for the matter dominated universe, i.e.
$H^2=k^2\rho_{m}/3$, solution of Eq. (\ref{eqdelta1}) yields
$\delta_{m}=a$. Hence, we introduce a new variable $g(a)$, namely
$g(a)=\delta_{m}/a$ which does not depend on $a$ during the matter
dominated era. Thus, the natural choice for the initial conditions
are $g(a_{\rm i})=1$ and $\frac{{\rm d}g}{{\rm d}\ln
a}\mid_{a=a_{\rm i}}=0$ \cite{Rui Zheng}. The initial moment should
be taken during the matter era, e.g., $a_{\rm i}=1/31$ (i.e.,
$z_{\rm i}=30$).

\begin{figure}
\includegraphics{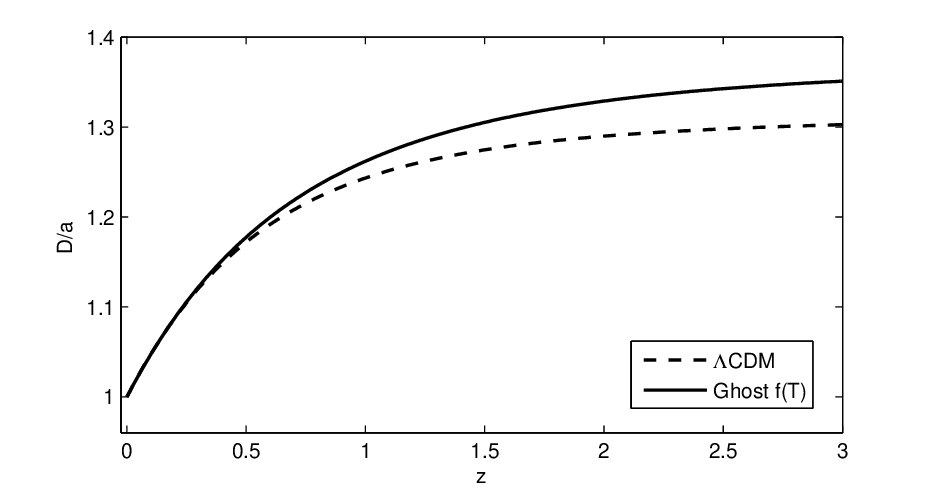}
      \vspace{4.5cm}
\caption[]{Linear growth function
$D=\frac{\delta_{m}}{\delta_{m_0}}$, normalized to today's value,
relative to its value in a pure-matter model ($D=a$), for the ghost
$f(T)$-gravity model and the $\Lambda$CDM model using the full data
sets.}
         \label{deltam}
   \end{figure}

\begin{figure}
\centering\includegraphics{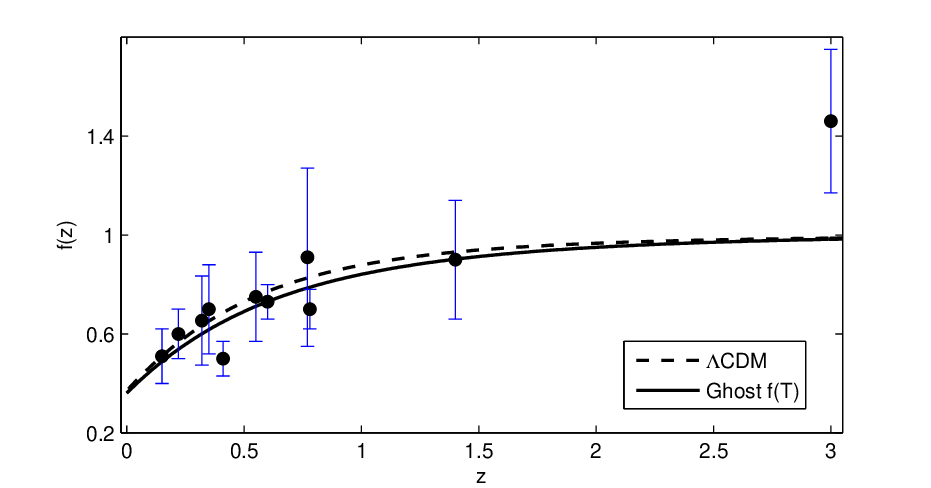}
      \vspace{4.5cm}
\caption[]{Evolution behaviors of the growth factor for the ghost
$f(T)$-gravity model and the $\Lambda$CDM model using the full data
sets.}
         \label{f}
  \end{figure}

In terms of new variable $g(a)$, Eq. (\ref{eqdelta1}) becomes
\begin{equation}
\frac{{\rm d}^2g}{{\rm
d}\ln{a^2}}+\left(4+\frac{\dot{H}}{{H}^{2}}\right)\frac{{\rm
d}g}{{\rm d}\ln{a}}
+\left(3+\frac{\dot{H}}{{H}^{2}}-\frac{4{\pi}G_{\rm
eff}\rho_{m}}{H^2}\right)g=0.\label{eqg1}
\end{equation}
From Eqs. (\ref{q1}), (\ref{q2}), (\ref{f(T)GDE}) and (\ref{beta})
one can obtain
\begin{equation}
\frac{\dot{H}}{H^2}=-\frac{3}{2}\left(\frac{2f_T-f/T}{f_T+2Tf_{TT}}\right)=-3\left[\frac{E+\Omega_{m_0}-1}{2E+\Omega_{m_0}-1}\right],\label{J1}
\end{equation}
where $E(z)=H(z)/H_0$ is given by Eq. (\ref{eqE}). Also with the
help of Eqs. (\ref{Hdot}) and (\ref{J1}) one can get
\begin{equation}
\frac{4{\pi}G_{\rm
eff}\rho_{m}}{H^2}=\frac{3}{2}\left(2-\frac{f}{Tf_T}\right)=3\left[\frac{E+\Omega_{m_0}-1}{2E+(\Omega_{m_0}-1)\ln
E}\right].\label{J2}
\end{equation}
In general, there is no analytical solution to Eq. (\ref{eqg1}), and
we need to solve it numerically. In Fig. \ref{deltam}, we plot the
evolutionary behavior of $g(a)$ normalized to today's value, with
the best fitting values of the QCD ghost $f(T)$-gravity model and
the $\Lambda$CDM model. Figure \ref{deltam} shows that $g(a)$ for
the ghost $f(T)$-gravity model like the $\Lambda$CDM decreases
during history of the universe. In Fig. \ref{f}, we also plot the
evolution of the growth factor $f(z)$ defined as \cite{Peebles}
\begin{equation}
f(z)=\frac{{\rm d}\ln\delta_{\rm m}}{{\rm d}\ln
a}=-(1+z)\,\frac{{\rm d}\ln\delta_{\rm m}}{{\rm d}z},
\end{equation}
with the best fitting values of the ghost $f(T)$-gravity model and
the $\Lambda$CDM model. The 11 data of the growth factor are
summarized in Table \ref{fdata}. Figure \ref{f} shows that the ghost
$f(T)$-gravity model can not be discriminated by the data, and both
this model and the $\Lambda$CDM model fit the data very well.
\section{Conclusions}

Here, we reconstructed a $f(T)$ model according to the QCD ghost DE
paradigm. In the framework of $f(T)$ modified teleparallel theory,
we considered a spatially flat FRW universe filled only with the
pressureless matter. Then, we obtained the time evolution of the
dark torsion scalar $T=-6H^2$ (or the Hubble parameter). We also
calculated the effective torsion EoS parameter of the ghost
$f(T)$-gravity model as well as the deceleration parameter of the
universe. Furthermore, we fitted the model with current
observational data, including SNeIa, CMB and BAO data. Using a
cosmographic analysis approach, we also checked the viability of our
model without the need of explicitly solving the field equations and
fitting the data. We further examined the validity of the GSL of
gravitational thermodynamics for the QCD ghost $f(T)$-gravity model.
Finally, we pointed out the growth of structure formation in our
model. Our results show the following.

(i) The condition $f/T\rightarrow 1$ is satisfied for our model at
high redshift ($|T|\rightarrow\infty$) which is compatible with the
primordial nucleosynthesis and CMB constraints.

(ii) The effective torsion EoS parameter $\omega_T$ varies from
$-0.5$ at early time to $-1$ at late time, which is similar to
freezing quintessence DE. For the present time we obtain
$\omega_{T_0}=-0.79$.

(iii) The variation of the deceleration parameter $q$ shows that the
universe transits from an early matter dominant epoch ($q=0.5$) to
the de Sitter era ($q=-1$) in the future, as expected. Also at the
near past $t/t_0=0.57$ we have a cosmic deceleration ($q>0$) to
acceleration ($q<0$) transition. The deceleration parameter
$q_0\simeq-0.4$ obtained at the present is compatible with the
recent observations.

(iv) The best fit value of the model parameter is
$\Omega_{m_0}=0.262_{-0.013}^{+0.013}(1\sigma)_{-0.025}^{+0.027}(2\sigma)$
for the full data sets including SNeIa+CMB+BAO. The minimal $\chi^2$
gives $\chi_{\rm min}^{2}=606.567$ with dof $=582$. The reduced
$\chi^2$ equals to $1.042$ which is acceptable. But $\chi_{\rm
min}^{2}$ is larger than the one for the $\Lambda$CDM model,
$\chi_{\Lambda\rm CDM}^{2}=562.531$, for the same data sets.

(v) Cosmographic analysis shows that the QCD ghost $f(T)$-gravity
model is favored by the observational data.

(vi) The GSL of gravitational thermodynamics holds for our $f(T)$
model throughout history of the universe.

(vii) The evolutionary behavior of the growth factor of matter
perturbation shows that the ghost $f(T)$-gravity model can not be
discriminated by the data, and both this model and the $\Lambda$CDM
model fit the data very well.

\subsection*{Acknowledgements}

The authors thank the reviewers for their valuable comments. The
works of K. Karami and Z. Safari have been supported financially by
Research Institute for Astronomy and Astrophysics of Maragha (RIAAM)
under research project No. 1/2782-41.


\clearpage
\clearpage
\clearpage
\begin{table}
  \centering
  \caption{The best fit value of $\Omega_{m_0}$ within the 68.3\% and 95.4\%
confidence intervals for each observational data set for the QCD
ghost $f(T)$-gravity model. The last column shows the best fit
result of the $\Lambda$CDM model using the full data sets for
comparison.}
  \label{best-fit}
  \begin{tabular}{ccccc}\noalign{\smallskip}
    \hline\hline
    Parameter & SN & SN+CMB & SN+CMB+BAO & $\Lambda$CDM\\\hline\\
    $\Omega_{m_0}$ & $0.179_{-0.014-0.028}^{+0.016+0.032}$ & $0.247_{-0.014-0.028}^{+0.014+0.030}$ &
    $0.262_{-0.013-0.025}^{+0.013+0.027}$ & $0.273_{-0.013-0.026}^{+0.014+0.028}$\\\\
        \hline\hline
  \end{tabular}
  \end{table}
\clearpage
\clearpage
  \begin{table}\small
\centering \caption{The observational data for the linear growth
rate $f_{\rm obs}(z)$.}
\begin{tabular}{lccccccccccc}\hline
$z$ & $0.15$ & $0.22$  & $0.32$ & $0.35$ & $0.41$ & $0.55$ & $0.60$
& $0.77$ & $0.78$ & $1.4$ & $3.0$
\\\hline
 $f_{\rm obs}$ &  $0.51$ &$0.60$ & $0.654$ & $0.70$ & $0.50$ & $0.75$ &
$0.73$  &$0.91$& $0.70$ & $0.90$ & $1.46$
\\\hline$1\sigma$ & $0.11$ &$0.10$ & $0.18$ & $0.18$ &
$0.07$ & $0.18$ & $0.07$ & $0.36$ & $0.08$ & $0.24$ & $0.29$
\\
\hline${\rm Ref.}$ & $\cite{Hawkins}$ &$\cite{Blake}$ &
$\cite{Reyes}$ & $\cite{Tegmarks}$ & $\cite{Blake}$ & $\cite{Ross}$
& $\cite{Blake}$ & $\cite{Guzzo}$ & $\cite{Blake}$ & $\cite{Angela}$
& $\cite{Donald}$\\\hline
\end{tabular}
\label{fdata}\\
\end{table}

\end{document}